\documentclass[12pt, letterpaper, onecolumn, peerreview]{IEEEtran}
\usepackage{color}
\usepackage{colordvi}
\usepackage{graphicx}
\usepackage{amsmath}
\usepackage{amssymb}
\usepackage{epic}

\newtheorem{theorem}{Theorem}[section]
\newtheorem{lemma}{Lemma}[section]

\begin{document}

\title{ \large \bf Anytime coding on the infinite bandwidth AWGN
  channel: \\ 
  A sequential semi-orthogonal optimal code \\}

\author{Anant Sahai\footnote{A.~Sahai is with Wireless Foundations in
    the Department of Electrical Engineering and Computer Sciences,
    U.C.~Berkeley. An early version of this paper was presented at the
    2005 CISS Conference in Baltimore, MD. } \\
  {\small sahai@eecs.berkeley.edu}}

\markboth{IEEE Transactions on Information Theory,~Vol.~??,
No.~??,~Month??~2005}{Shell \MakeLowercase{\textit{et al.}}: Bare
Demo of IEEEtran.cls for Journals}

%

\maketitle

\begin{abstract}
  It is well known that orthogonal coding can be used to approach the
  Shannon capacity of the power-constrained AWGN channel without a
  bandwidth constraint.  This correspondence describes a
  semi-orthogonal variation of pulse position modulation that is
  sequential in nature --- bits can be ``streamed across'' without
  having to buffer up blocks of bits at the transmitter. ML decoding
  results in an exponentially small probability of error as a function
  of tolerated receiver delay and thus eventually a zero probability
  of error on every transmitted bit. In the high-rate regime, a
  matching upper bound is given on the delay error exponent. We
  close with some comments on the case with feedback and the
  connections to the capacity per unit cost problem.
\end{abstract}


\IEEEpeerreviewmaketitle

\section{Introduction} \label{sec:intro}

Shannon's capacity theorem is arguably the greatest accomplishment in
communications theory. Unfortunately, the random coding proof is
non-constructive in that does not give a construction for any explicit
code. This has led to the proverb: ``Almost all codes are "good" codes
except for the the ones that we can think of.''\cite{WolfLecture} However,
there is a channel for which explicit non-random constructions for
capacity-achieving codes exist: the continuous-time AWGN channel with
an input power constraint and no bandwidth constraint. (See, e.g.
\cite{gallager}) For such channels, it is further known that
orthogonal signaling can be used to achieve data rates arbitrarily
close to capacity. Orthogonality of the codewords plays the role of
codeword independence in the discrete case. Recently, Liu and
Viswanath have shown in \cite{LiuViswanath} how to extend orthogonal
coding to the ``writing on dirty paper'' scenario while continuing to
preserve the interference-free infinite bandwidth error exponents.
This correspondence extends orthogonal coding in a different direction
to deal with ``streaming data.''

\subsection{Model and block-coding review}
The noise process is modeled as white with intensity $\frac{N_0}{2}$.
The capacity of the channel is most naturally expressed in terms of
energy per bit and is given by:
\begin{equation} \label{eqn:awgncapacityinenergy}
E_b > N_0 \ln 2
\end{equation}
which means that reliable communication is possible if the normalized
energy per-bit exceeds $\ln 2$. When viewed in terms of bits per unit
time, it means that reliable communication requires:
\begin{equation} \label{eqn:awgncapacityintime}
R < C_\infty = \frac{P}{N_0} \log_2 e
\end{equation}
where $P$ represents the allowed power per unit time. 

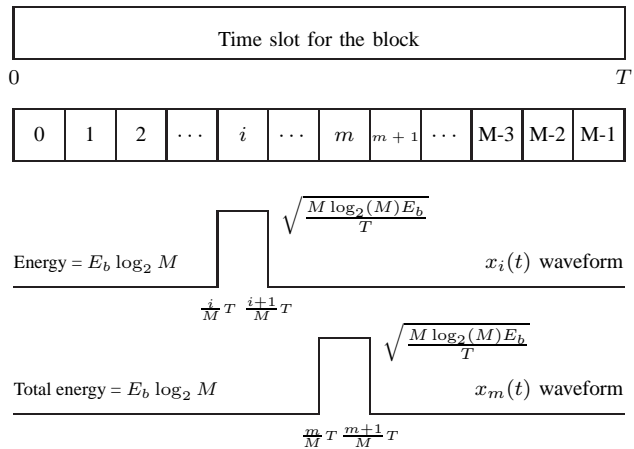
\begin{figure} 
\begin{center}
\setlength{\unitlength}{2100sp}%
\begingroup\makeatletter\ifx\SetFigFont\undefined%
\gdef\SetFigFont#1#2#3#4#5{%
  \reset@font\fontsize{#1}{#2pt}%
  \fontfamily{#3}\fontseries{#4}\fontshape{#5}%
  \selectfont}%
\fi\endgroup%
\begin{picture}(7224,5170)(1189,-4619)
\thinlines
\put(1201,-61){\framebox(7200,600){}}%
\put(1201,-1261){\framebox(7200,600){}}
\put(6601,-661){\line( 0,-1){600}}
\put(7201,-661){\line( 0,-1){600}}
\put(7801,-661){\line( 0,-1){600}}
\put(8401,-661){\line( 0,-1){600}}
\put(4201,-661){\line( 0,-1){600}}
\put(4801,-661){\line( 0,-1){600}}
\put(5401,-661){\line( 0,-1){600}}
\put(6001,-661){\line( 0,-1){600}}
\put(1801,-661){\line( 0,-1){600}}
\put(2401,-661){\line( 0,-1){600}}
\put(3001,-661){\line( 0,-1){600}}
\put(3601,-661){\line( 0,-1){600}}
\put(1201,-2761){\line( 1, 0){2400}}
\put(3601,-2761){\line( 0, 1){900}}
\put(3601,-1861){\line( 1, 0){600}}
\put(4201,-1861){\line( 0,-1){900}}
\put(4201,-2761){\line( 1, 0){4200}}
\put(1201,-4261){\line( 1, 0){3600}}
\put(4801,-4261){\line( 0, 1){900}}
\put(4801,-3361){\line( 1, 0){600}}
\put(5401,-3361){\line( 0,-1){900}}
\put(5401,-4261){\line( 1, 0){3000}}
\put(4801, 89){\makebox(0,0)[b]{\smash{\SetFigFont{8}{10}{\rmdefault}{\mddefault}{\updefault}{\color[rgb]{0,0,0}Time slot for the block}%
}}}
\put(1501,-1036){\makebox(0,0)[b]{\smash{\SetFigFont{8}{10}{\rmdefault}{\mddefault}{\updefault}{\color[rgb]{0,0,0}0}%
}}}
\put(2101,-1036){\makebox(0,0)[b]{\smash{\SetFigFont{8}{10}{\rmdefault}{\mddefault}{\updefault}{\color[rgb]{0,0,0}1}%
}}}
\put(2701,-1036){\makebox(0,0)[b]{\smash{\SetFigFont{8}{10}{\rmdefault}{\mddefault}{\updefault}{\color[rgb]{0,0,0}2}%
}}}
\put(8101,-1036){\makebox(0,0)[b]{\smash{\SetFigFont{8}{10}{\rmdefault}{\mddefault}{\updefault}{\color[rgb]{0,0,0}M-1}%
}}}
\put(7501,-1036){\makebox(0,0)[b]{\smash{\SetFigFont{8}{10}{\rmdefault}{\mddefault}{\updefault}{\color[rgb]{0,0,0}M-2}%
}}}
\put(6901,-1036){\makebox(0,0)[b]{\smash{\SetFigFont{8}{10}{\rmdefault}{\mddefault}{\updefault}{\color[rgb]{0,0,0}M-3}%
}}}
\put(6301,-1036){\makebox(0,0)[b]{\smash{\SetFigFont{8}{10}{\rmdefault}{\mddefault}{\updefault}{\color[rgb]{0,0,0}$\cdots$}%
}}}
\put(3301,-1036){\makebox(0,0)[b]{\smash{\SetFigFont{8}{10}{\rmdefault}{\mddefault}{\updefault}{\color[rgb]{0,0,0}$\cdots$}%
}}}
\put(3901,-1036){\makebox(0,0)[b]{\smash{\SetFigFont{8}{10}{\rmdefault}{\mddefault}{\updefault}{\color[rgb]{0,0,0}$i$}%
}}}
\put(4501,-1036){\makebox(0,0)[b]{\smash{\SetFigFont{8}{10}{\rmdefault}{\mddefault}{\updefault}{\color[rgb]{0,0,0}$\cdots$}%
}}}
\put(5101,-1036){\makebox(0,0)[b]{\smash{\SetFigFont{8}{10}{\rmdefault}{\mddefault}{\updefault}{\color[rgb]{0,0,0}$m$}%
}}}
\put(5701,-1036){\makebox(0,0)[b]{\smash{\SetFigFont{5}{10}{\rmdefault}{\mddefault}{\updefault}{\color[rgb]{0,0,0}$m+1$}%
}}}
\put(8401,-2536){\makebox(0,0)[rb]{\smash{\SetFigFont{8}{10}{\rmdefault}{\mddefault}{\updefault}{\color[rgb]{0,0,0}$x_i(t)$ waveform}%
}}}
\put(8401,-4036){\makebox(0,0)[rb]{\smash{\SetFigFont{8}{10}{\rmdefault}{\mddefault}{\updefault}{\color[rgb]{0,0,0}$x_m(t)$ waveform}%
}}}
\put(5401,-4561){\makebox(0,0)[b]{\smash{\SetFigFont{5}{10}{\rmdefault}{\mddefault}{\updefault}{\color[rgb]{0,0,0}$\frac{m+1}{M}T$}%
}}}
\put(4801,-4561){\makebox(0,0)[b]{\smash{\SetFigFont{5}{10}{\rmdefault}{\mddefault}{\updefault}{\color[rgb]{0,0,0}$\frac{m}{M}T$}%
}}}
\put(3601,-3061){\makebox(0,0)[b]{\smash{\SetFigFont{5}{10}{\rmdefault}{\mddefault}{\updefault}{\color[rgb]{0,0,0}$\frac{i}{M}T$}%
}}}
\put(4201,-3061){\makebox(0,0)[b]{\smash{\SetFigFont{5}{10}{\rmdefault}{\mddefault}{\updefault}{\color[rgb]{0,0,0}$\frac{i+1}{M}T$}%
}}}
\put(1201,-361){\makebox(0,0)[b]{\smash{\SetFigFont{8}{10}{\rmdefault}{\mddefault}{\updefault}{\color[rgb]{0,0,0}$0$}%
}}}
\put(8401,-361){\makebox(0,0)[b]{\smash{\SetFigFont{8}{10}{\rmdefault}{\mddefault}{\updefault}{\color[rgb]{0,0,0}$T$}%
}}}
\put(4351,-2011){\makebox(0,0)[lb]{\smash{\SetFigFont{8}{10}{\rmdefault}{\mddefault}{\updefault}{\color[rgb]{0,0,0}$\sqrt{\frac{M \log_2(M) E_b}{T}}$}%
}}}
\put(5551,-3511){\makebox(0,0)[lb]{\smash{\SetFigFont{8}{10}{\rmdefault}{\mddefault}{\updefault}{\color[rgb]{0,0,0}$\sqrt{\frac{M \log_2(M) E_b}{T}}$}%
}}}
\put(1201,-2536){\makebox(0,0)[lb]{\smash{\SetFigFont{7}{10}{\rmdefault}{\mddefault}{\updefault}{\color[rgb]{0,0,0}Energy = $E_b \log_2 M$}%
}}}
\put(1201,-4036){\makebox(0,0)[lb]{\smash{\SetFigFont{7}{10}{\rmdefault}{\mddefault}{\updefault}{\color[rgb]{0,0,0}Total energy = $E_b \log_2 M$}%
}}}
\end{picture}
\end{center}
\caption{Pulse position modulation illustrated in the block coding
  framework.}
\label{fig:blockorthogonal}
\end{figure}

One orthogonal signaling scheme is pulse position modulation (PPM) as
depicted in Figure~\ref{fig:blockorthogonal}. Suppose there are $M$
possible messages to distinguish during one time slot of duration $T$.
To communicate message $0 \leq m \leq M-1$, the transmitter sends out
a burst of its allocated transmit power during the time slot
$[\frac{m}{M}T, \frac{m+1}{M}T]$ and is silent during the rest of the
time slots. Since the time-slots are disjoint, the waveforms $x_m(t)$
corresponding to different messages $m$ are necessarily orthogonal
over the interval $[0,T]$.

\begin{figure}
\begin{center}
\includegraphics[width=4in]{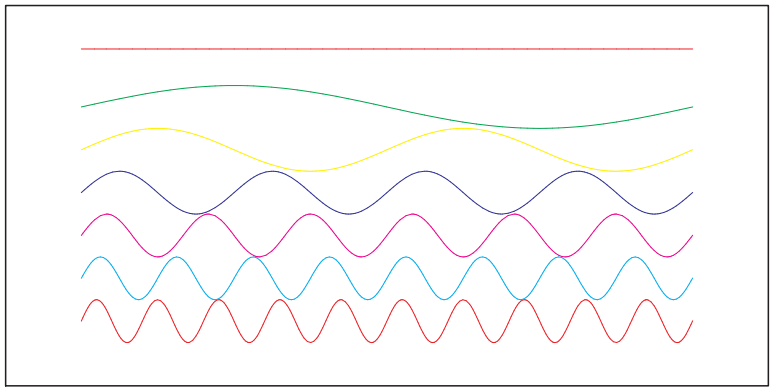}
\end{center}
\caption{Another set of orthogonal waveforms: sinusoids of
  differing frequencies.}
\label{fig:blockwaveforms}
\end{figure}

Another set of orthogonal waveforms, more suitable for channels with
amplitude-limits at the input, is depicted in
Figure~\ref{fig:blockwaveforms}. These are sinusoids of different
frequencies. If there are only amplitude-constraints but no binding
power constraints, then ``bang-bang'' versions (square-waves) can also
give orthogonal waveforms that have maximal energy given the amplitude
constraint. 

The receiver can do simple maximum-likelihood detection by having a
bank of $M$ matched-filters that correlate the received signal $Y(t)$
with the $M$ orthogonal waveforms. The result of this correlation can be
considered as $Z_i$ for message $i$ and due to the AWGN assumption,
can be modeled as:
\begin{equation} \label{eqn:orthogonalmodel}
Z_i = \left\{ \begin{array}{ll} 
              N_i & \mbox{if }i\neq m \\
              \sqrt{2 E_b \log_2 M } + N_m & \mbox{if }i=m 
	      \end{array} \right.
\end{equation}
where the $N_i$ are iid~standard zero-mean unit-variance\footnote{The
$2 E_b$ term is there to achieve this normalization of the noise.}
Gaussian random variables, $E_b$ represents the normalized energy per
bit, and $m$ represents the true message sent. $\log_2 M$ represents
the number of bits to be communicated.

In this case, ML decoding consists of finding the waveform with the
highest $Z_i$. Classical analysis of this channel (see,
e.g.~\cite{gallager}) shows that the orthogonal code with
maximum-likelihood (ML) decoding has a probability of block error that
goes to zero exponentially with block duration:
\begin{equation} \label{eqn:proberrorblock}
P_e \leq K e^{-T E_{orth}(R)}
\end{equation}
where $K$ is a rate dependent constant and 
\begin{equation} \label{eqn:orthogonalexponent}
E_{orth}(R) = \left\{ \begin{array}{rl} 
                      (\frac{C_\infty}{2} - R)\ln 2 & \mbox{if }
                      0 \leq R \leq \frac{C_\infty}{4} \\
                      (\sqrt{C_\infty} - \sqrt{R})^2\ln 2 & \mbox{if }
                      \frac{C_\infty}{4} < R < C_{\infty} \\ 
                      0 & \mbox{otherwise} \end{array}\right.
\end{equation}

Wyner showed in \cite{WynerAWGN} that $E_{orth}(R)$ is also the best
possible error exponent for this block-coding problem.

\subsection{Non-block coding}
For a situation in which bits arrive from the source at regular
intervals, the traditional view involves buffering up a block of
$\log_2 M$ bits, and then sending out the block of data using
orthogonal signaling while waiting for the next block of data bits to
arrive at the encoder. If there is only an energy constraint, then
because there is no bandwidth constraint, the duration of signaling
can be made as small as desired and hence essentially all the
end-to-end delay can be attributed to the buffering at the
encoder.\footnote{This is different in the traditional picture as
  applied to a DMC or any other finite degree-of-freedom channel. In
  that picture, $T$ must be of a certain length and hence the delay is
  essentially split between the encoder and the decoder resulting in
  an end-to-end delay of around $2N - 1$ channel uses for a block-code
  of size $N$.}

\begin{figure}
\begin{center}
\setlength{\unitlength}{2300sp}%
\begingroup\makeatletter\ifx\SetFigFont\undefined%
\gdef\SetFigFont#1#2#3#4#5{%
  \reset@font\fontsize{#1}{#2pt}%
  \fontfamily{#3}\fontseries{#4}\fontshape{#5}%
  \selectfont}%
\fi\endgroup%
\begin{picture}(6044,1834)(1179,-2483)
\thinlines
{\color[rgb]{0,0,0}\put(4801,-661){\line( 0,-1){600}}
}%
{\color[rgb]{0,0,0}\put(6001,-661){\line( 0,-1){600}}
}%
{\color[rgb]{0,0,0}\put(2401,-661){\line( 0,-1){600}}
}%
{\color[rgb]{0,0,0}\put(3601,-661){\line( 0,-1){600}}
}%
{\color[rgb]{0,0,0}\put(1201,-1261){\framebox(6000,600){}}
}%
\thicklines
{\color[rgb]{0,0,0}\put(1201,-2461){\line( 1, 0){300}}
\put(1501,-2461){\line( 0, 1){600}}
\put(1501,-1861){\line( 1, 0){300}}
\put(1801,-1861){\line( 0,-1){600}}
\put(1801,-2461){\line( 1, 0){1200}}
\put(3001,-2461){\line( 0, 1){600}}
\put(3001,-1861){\line( 1, 0){300}}
\put(3301,-1861){\line( 0,-1){600}}
\put(3301,-2461){\line( 1, 0){300}}
}%
{\color[rgb]{0,0,0}\put(4801,-2461){\line( 0, 1){600}}
\put(4801,-1861){\line( 1, 0){300}}
\put(5101,-1861){\line( 0,-1){600}}
\put(5101,-2461){\line( 1, 0){1500}}
\put(6601,-2461){\line( 0, 1){600}}
\put(6601,-1861){\line( 1, 0){300}}
\put(6901,-1861){\line( 0,-1){600}}
\put(6901,-2461){\line( 1, 0){300}}
}%
\thinlines
{\color[rgb]{0,0,0}\put(3601,-2461){\line( 1, 0){ 75}}
\put(3676,-2461){\line( 0, 1){600}}
\put(3676,-1861){\line( 1, 0){ 75}}
\put(3751,-1861){\line( 0,-1){600}}
\put(3751,-2461){\line( 1, 0){150}}
\put(3901,-2461){\line( 0, 1){600}}
\put(3901,-1861){\line( 1, 0){ 75}}
\put(3976,-1861){\line( 0,-1){600}}
\put(3976,-2461){\line( 1, 0){225}}
\put(4201,-2461){\line( 0, 1){600}}
\put(4201,-1861){\line( 1, 0){ 75}}
\put(4276,-1861){\line( 0,-1){600}}
\put(4276,-2461){\line( 1, 0){300}}
\put(4576,-2461){\line( 0, 1){600}}
\put(4576,-1861){\line( 1, 0){ 75}}
\put(4651,-1861){\line( 0,-1){600}}
\put(4651,-2461){\line( 1, 0){150}}
}%
\put(1201,-1561){\makebox(0,0)[b]{\smash{\SetFigFont{8}{10}{\rmdefault}{\mddefault}{\updefault}{\color[rgb]{0,0,0}$0$}%
}}}
\put(2401,-1561){\makebox(0,0)[b]{\smash{\SetFigFont{8}{10}{\rmdefault}{\mddefault}{\updefault}{\color[rgb]{0,0,0}$T$}%
}}}
\put(1801,-1036){\makebox(0,0)[b]{\smash{\SetFigFont{8}{10}{\rmdefault}{\mddefault}{\updefault}{\color[rgb]{0,0,0}block 0}%
}}}
\put(4201,-1036){\makebox(0,0)[b]{\smash{\SetFigFont{8}{10}{\rmdefault}{\mddefault}{\updefault}{\color[rgb]{0,0,0}$\cdots$}%
}}}
\put(3001,-1036){\makebox(0,0)[b]{\smash{\SetFigFont{8}{10}{\rmdefault}{\mddefault}{\updefault}{\color[rgb]{0,0,0}block 1}%
}}}
\put(5401,-1036){\makebox(0,0)[b]{\smash{\SetFigFont{8}{10}{\rmdefault}{\mddefault}{\updefault}{\color[rgb]{0,0,0}$block k$}%
}}}
\put(6601,-1036){\makebox(0,0)[b]{\smash{\SetFigFont{8}{10}{\rmdefault}{\mddefault}{\updefault}{\color[rgb]{0,0,0}$\cdots$}%
}}}
\put(3601,-1561){\makebox(0,0)[b]{\smash{\SetFigFont{8}{10}{\rmdefault}{\mddefault}{\updefault}{\color[rgb]{0,0,0}$2T$}%
}}}
\put(4801,-1561){\makebox(0,0)[b]{\smash{\SetFigFont{8}{10}{\rmdefault}{\mddefault}{\updefault}{\color[rgb]{0,0,0}$kT$}%
}}}
\put(6001,-1561){\makebox(0,0)[b]{\smash{\SetFigFont{8}{10}{\rmdefault}{\mddefault}{\updefault}{\color[rgb]{0,0,0}$(k+1)T$}%
}}}
\end{picture}
\end{center}
\caption{One block after another where each block uses an orthogonal
  code.}
\label{fig:many_blocks}
\end{figure}
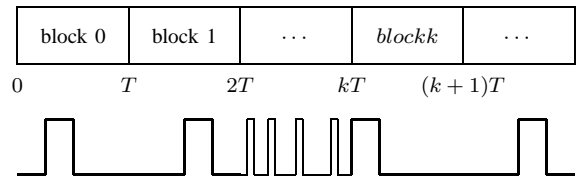

For DMCs and finite-bandwidth channels, convolutional codes and
tree-codes provide an alternative to using block-codes. These codes
have no buffering delay at the encoder and instead delay is incurred
at the receiver. For DMCs, in principle random convolutional or tree
codes eventually allow every bit to be decoded correctly, with a
probability of error that converges to zero exponentially in delay at
a rate given by the random coding error exponent \cite{ForneyML}.
Furthermore, at high rates for symmetric channels, it is known that no
faster convergence is possible without
feedback \cite{PinskerNoFeedback, OurUpperBoundPaper}. 

There are also efficient decoding algorithms that work very well at
low rates \cite{JelinekSequential, ForneySeq} assuming oracle access
to the code. Random convolutional codes can also be implemented using
a computational cost that is linear in the constraint-length if the
constraint is bounded, and linearly increasing with time if the
constraint-length is infinite. If perfect tentative decision feedback
is available, \cite{OurUpperBoundPaper} also gives a scheme that gives
infinite-constraint-length performance using only bounded expected
computation per unit time. 

To our knowledge, the infinite-bandwidth Gaussian counterparts to
these ideas have not been developed in the literature. While clearly
impractical, the codes in this paper have pedagogic value since they
are explicit and so in many ways simpler than the random coding
constructions for DMCs.

\section{The semi-orthogonal code}

\subsection{Motivation: zero-rate coding by repetition} \label{sec:zerorate}

Consider a binary symmetric channel. The repetition code is an
excellent code at zero-rate. To achieve a target reliability, just
repeat a bit as many times as needed. To achieve perfect reliability,
just repeat it infinitely often. Now, suppose that bits were arriving
at the encoder regularly, but we neither cared about the delay in
decoding them at the decoder nor about any increase in this delay with
bit position. One strategy to communicate reliably would be as
follows:
\begin{enumerate}
 \item Initialize $i=1$
 \item Output every bit $B_1^i$
 \item Increment $i$
 \item goto step 2.
\end{enumerate}

This would result in an output stream $B_1; B_1, B_2; B_1, B_2, B_3;
B_1, B_2, B_3, B_4; \ldots$ where the semicolons are used to denote
the points of time at which we start repeating bits again and the
commas mark off the channel uses. It is clear that if the decoder
waits long enough, it will get enough repetitions of any individual
bit to achieve its target reliability.

While this code has zero rate and has increasing required delay with
time, it is possible to modify this scheme for use on the infinite
bandwidth AWGN channel with finite rate, finite energy per bit, and
fixed delays.

\subsection{Repeated/refined pulse position modulation}

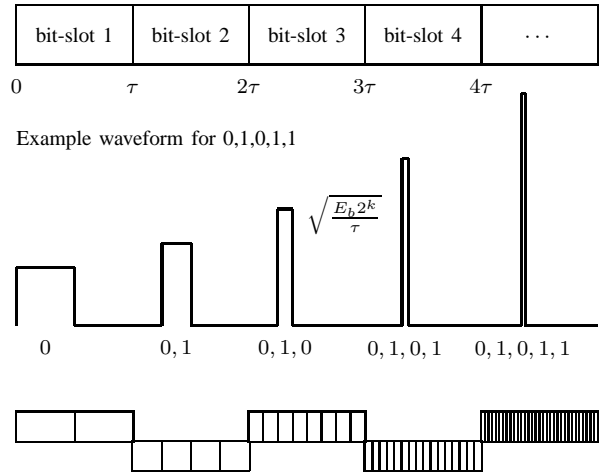
\begin{figure}
\begin{center}
\setlength{\unitlength}{2400sp}%
\begingroup\makeatletter\ifx\SetFigFont\undefined%
\gdef\SetFigFont#1#2#3#4#5{%
  \reset@font\fontsize{#1}{#2pt}%
  \fontfamily{#3}\fontseries{#4}\fontshape{#5}%
  \selectfont}%
\fi\endgroup%
\begin{picture}(6044,4824)(1179,-5473)
\thinlines
{\color[rgb]{0,0,0}\put(1201,-5161){\framebox(1200,300){}}
}%
{\color[rgb]{0,0,0}\put(1801,-4861){\line( 0,-1){300}}
}%
{\color[rgb]{0,0,0}\put(2401,-5461){\framebox(1200,300){}}
}%
{\color[rgb]{0,0,0}\put(3001,-5161){\line( 0,-1){300}}
}%
{\color[rgb]{0,0,0}\put(2701,-5161){\line( 0,-1){300}}
}%
{\color[rgb]{0,0,0}\put(3301,-5161){\line( 0,-1){300}}
}%
{\color[rgb]{0,0,0}\put(3601,-5161){\framebox(1200,300){}}
}%
{\color[rgb]{0,0,0}\put(4201,-4861){\line( 0,-1){300}}
}%
{\color[rgb]{0,0,0}\put(3901,-4861){\line( 0,-1){300}}
}%
{\color[rgb]{0,0,0}\put(4501,-4861){\line( 0,-1){300}}
}%
{\color[rgb]{0,0,0}\put(3751,-4861){\line( 0,-1){300}}
}%
{\color[rgb]{0,0,0}\put(4051,-4861){\line( 0,-1){300}}
}%
{\color[rgb]{0,0,0}\put(4351,-4861){\line( 0,-1){300}}
}%
{\color[rgb]{0,0,0}\put(4651,-4861){\line( 0,-1){300}}
}%
{\color[rgb]{0,0,0}\put(5401,-5161){\line( 0,-1){300}}
}%
{\color[rgb]{0,0,0}\put(5101,-5161){\line( 0,-1){300}}
}%
{\color[rgb]{0,0,0}\put(5701,-5161){\line( 0,-1){300}}
}%
{\color[rgb]{0,0,0}\put(4951,-5161){\line( 0,-1){300}}
}%
{\color[rgb]{0,0,0}\put(5251,-5161){\line( 0,-1){300}}
}%
{\color[rgb]{0,0,0}\put(5551,-5161){\line( 0,-1){300}}
}%
{\color[rgb]{0,0,0}\put(5851,-5161){\line( 0,-1){300}}
}%
{\color[rgb]{0,0,0}\put(4876,-5161){\line( 0,-1){300}}
}%
{\color[rgb]{0,0,0}\put(5026,-5161){\line( 0,-1){300}}
}%
{\color[rgb]{0,0,0}\put(5176,-5161){\line( 0,-1){300}}
}%
{\color[rgb]{0,0,0}\put(5401,-5161){\line( 0,-1){300}}
}%
{\color[rgb]{0,0,0}\put(5326,-5161){\line( 0,-1){300}}
}%
{\color[rgb]{0,0,0}\put(5476,-5161){\line( 0,-1){300}}
}%
{\color[rgb]{0,0,0}\put(5626,-5161){\line( 0,-1){300}}
}%
{\color[rgb]{0,0,0}\put(5776,-5161){\line( 0,-1){300}}
}%
{\color[rgb]{0,0,0}\put(4801,-5461){\framebox(1200,300){}}
}%
{\color[rgb]{0,0,0}\put(5926,-5161){\line( 0,-1){300}}
}%
{\color[rgb]{0,0,0}\put(6638,-4853){\line( 0,-1){300}}
}%
{\color[rgb]{0,0,0}\put(6338,-4853){\line( 0,-1){300}}
}%
{\color[rgb]{0,0,0}\put(6938,-4853){\line( 0,-1){300}}
}%
{\color[rgb]{0,0,0}\put(6188,-4853){\line( 0,-1){300}}
}%
{\color[rgb]{0,0,0}\put(6488,-4853){\line( 0,-1){300}}
}%
{\color[rgb]{0,0,0}\put(6788,-4853){\line( 0,-1){300}}
}%
{\color[rgb]{0,0,0}\put(7088,-4853){\line( 0,-1){300}}
}%
{\color[rgb]{0,0,0}\put(6113,-4853){\line( 0,-1){300}}
}%
{\color[rgb]{0,0,0}\put(6263,-4853){\line( 0,-1){300}}
}%
{\color[rgb]{0,0,0}\put(6413,-4853){\line( 0,-1){300}}
}%
{\color[rgb]{0,0,0}\put(6638,-4853){\line( 0,-1){300}}
}%
{\color[rgb]{0,0,0}\put(6563,-4853){\line( 0,-1){300}}
}%
{\color[rgb]{0,0,0}\put(6713,-4853){\line( 0,-1){300}}
}%
{\color[rgb]{0,0,0}\put(6863,-4853){\line( 0,-1){300}}
}%
{\color[rgb]{0,0,0}\put(7013,-4853){\line( 0,-1){300}}
}%
{\color[rgb]{0,0,0}\put(7163,-4853){\line( 0,-1){300}}
}%
{\color[rgb]{0,0,0}\put(6601,-4861){\line( 0,-1){300}}
}%
{\color[rgb]{0,0,0}\put(6301,-4861){\line( 0,-1){300}}
}%
{\color[rgb]{0,0,0}\put(6901,-4861){\line( 0,-1){300}}
}%
{\color[rgb]{0,0,0}\put(6151,-4861){\line( 0,-1){300}}
}%
{\color[rgb]{0,0,0}\put(6451,-4861){\line( 0,-1){300}}
}%
{\color[rgb]{0,0,0}\put(6751,-4861){\line( 0,-1){300}}
}%
{\color[rgb]{0,0,0}\put(7051,-4861){\line( 0,-1){300}}
}%
{\color[rgb]{0,0,0}\put(6076,-4861){\line( 0,-1){300}}
}%
{\color[rgb]{0,0,0}\put(6226,-4861){\line( 0,-1){300}}
}%
{\color[rgb]{0,0,0}\put(6376,-4861){\line( 0,-1){300}}
}%
{\color[rgb]{0,0,0}\put(6601,-4861){\line( 0,-1){300}}
}%
{\color[rgb]{0,0,0}\put(6526,-4861){\line( 0,-1){300}}
}%
{\color[rgb]{0,0,0}\put(6676,-4861){\line( 0,-1){300}}
}%
{\color[rgb]{0,0,0}\put(6826,-4861){\line( 0,-1){300}}
}%
{\color[rgb]{0,0,0}\put(6976,-4861){\line( 0,-1){300}}
}%
{\color[rgb]{0,0,0}\put(6001,-5161){\framebox(1200,300){}}
}%
{\color[rgb]{0,0,0}\put(7126,-4861){\line( 0,-1){300}}
}%
{\color[rgb]{0,0,0}\put(6038,-4869){\line( 0,-1){300}}
}%
{\color[rgb]{0,0,0}\put(4801,-661){\line( 0,-1){600}}
}%
{\color[rgb]{0,0,0}\put(6001,-661){\line( 0,-1){600}}
}%
{\color[rgb]{0,0,0}\put(2401,-661){\line( 0,-1){600}}
}%
{\color[rgb]{0,0,0}\put(3601,-661){\line( 0,-1){600}}
}%
{\color[rgb]{0,0,0}\put(1201,-1261){\framebox(6000,600){}}
}%
\thicklines
{\color[rgb]{0,0,0}\put(1201,-3961){\line( 0, 1){600}}
\put(1201,-3361){\line( 1, 0){600}}
\put(1801,-3361){\line( 0,-1){600}}
\put(1801,-3961){\line( 1, 0){900}}
\put(2701,-3961){\line( 0, 1){844}}
\put(2700,-3117){\line( 1, 0){304}}
\put(3004,-3117){\line( 0,-1){844}}
\put(3001,-3961){\line( 1, 0){900}}
\put(3901,-3961){\line( 0, 1){1200}}
\put(3901,-2761){\line( 1, 0){150}}
\put(4051,-2761){\line( 0,-1){1200}}
\put(4051,-3961){\line( 1, 0){1132}}
\put(5183,-3960){\line( 0, 1){1725}}
\put(5183,-2235){\line( 1, 0){ 68}}
\put(5251,-2236){\line( 0,-1){1725}}
\put(5251,-3961){\line( 1, 0){1169}}
\put(6420,-3960){\line( 0, 1){2396}}
\put(6417,-1564){\line( 1, 0){ 37}}
\put(6454,-1564){\line( 0,-1){2397}}
\put(6451,-3961){\line( 1, 0){750}}
}%
\put(1201,-1561){\makebox(0,0)[b]{\smash{\SetFigFont{8}{10}{\rmdefault}{\mddefault}{\updefault}{\color[rgb]{0,0,0}$0$}%
}}}
\put(6601,-1036){\makebox(0,0)[b]{\smash{\SetFigFont{8}{10}{\rmdefault}{\mddefault}{\updefault}{\color[rgb]{0,0,0}$\cdots$}%
}}}
\put(2401,-1561){\makebox(0,0)[b]{\smash{\SetFigFont{8}{10}{\rmdefault}{\mddefault}{\updefault}{\color[rgb]{0,0,0}$\tau$}%
}}}
\put(3601,-1561){\makebox(0,0)[b]{\smash{\SetFigFont{8}{10}{\rmdefault}{\mddefault}{\updefault}{\color[rgb]{0,0,0}$2\tau$}%
}}}
\put(4801,-1561){\makebox(0,0)[b]{\smash{\SetFigFont{8}{10}{\rmdefault}{\mddefault}{\updefault}{\color[rgb]{0,0,0}$3\tau$}%
}}}
\put(6001,-1561){\makebox(0,0)[b]{\smash{\SetFigFont{8}{10}{\rmdefault}{\mddefault}{\updefault}{\color[rgb]{0,0,0}$4\tau$}%
}}}
\put(1201,-2100){\makebox(0,0)[lb]{\smash{\SetFigFont{8}{10}{\rmdefault}{\mddefault}{\updefault}{\color[rgb]{0,0,0}Example waveform for 0,1,0,1,1}%
}}}
\put(4201,-2911){\makebox(0,0)[lb]{\smash{\SetFigFont{8}{10}{\rmdefault}{\mddefault}{\updefault}{\color[rgb]{0,0,0}$\sqrt{\frac{E_b 2^k}{\tau}}$}%
}}}
\put(1501,-4261){\makebox(0,0)[b]{\smash{\SetFigFont{8}{10}{\rmdefault}{\mddefault}{\updefault}{\color[rgb]{0,0,0}$0$}%
}}}
\put(2851,-4261){\makebox(0,0)[b]{\smash{\SetFigFont{8}{10}{\rmdefault}{\mddefault}{\updefault}{\color[rgb]{0,0,0}$0,1$}%
}}}
\put(3976,-4261){\makebox(0,0)[b]{\smash{\SetFigFont{8}{10}{\rmdefault}{\mddefault}{\updefault}{\color[rgb]{0,0,0}$0,1,0$}%
}}}
\put(5212,-4265){\makebox(0,0)[b]{\smash{\SetFigFont{8}{10}{\rmdefault}{\mddefault}{\updefault}{\color[rgb]{0,0,0}$0,1,0,1$}%
}}}
\put(6430,-4265){\makebox(0,0)[b]{\smash{\SetFigFont{8}{10}{\rmdefault}{\mddefault}{\updefault}{\color[rgb]{0,0,0}$0,1,0,1,1$}%
}}}
\put(1801,-1036){\makebox(0,0)[b]{\smash{\SetFigFont{8}{10}{\rmdefault}{\mddefault}{\updefault}{\color[rgb]{0,0,0}bit-slot 1}%
}}}
\put(3001,-1036){\makebox(0,0)[b]{\smash{\SetFigFont{8}{10}{\rmdefault}{\mddefault}{\updefault}{\color[rgb]{0,0,0}bit-slot 2}%
}}}
\put(4201,-1036){\makebox(0,0)[b]{\smash{\SetFigFont{8}{10}{\rmdefault}{\mddefault}{\updefault}{\color[rgb]{0,0,0}bit-slot 3}%
}}}
\put(5401,-1036){\makebox(0,0)[b]{\smash{\SetFigFont{8}{10}{\rmdefault}{\mddefault}{\updefault}{\color[rgb]{0,0,0}bit-slot 4}%
}}}
\end{picture}
\end{center}
\caption{Repeated pulse position modulation illustrated. The time
slots are on the top and the possible sub-slots are on the bottom.}
\label{fig:many_bits}
\end{figure}

Suppose that bits are arriving every $\tau$ seconds and the encoder is
allowed $E_b$ energy per bit. Rather than waiting to build up a buffer
of $\log_2 M$ bits and then signaling, suppose the encoder instead
spent the $E_b$ energy immediately and used it to ``repeat'' the value
of every bit received so far as in the scheme of
Section~\ref{sec:zerorate}.  The scheme is illustrated in
Figure~\ref{fig:many_bits}. It ``refines'' the information as time
goes one.

\begin{lemma} \label{lem:semiorthogonal}
The repeated pulse position modulation code is semi-orthogonal.

If $x$ is the waveform corresponding to the bitstream $b$, and
$x'$ is the waveform corresponding to the bitstream $b'$, then
$x([(j-1)\tau, T]$ is orthogonal to $x'([(j-1)\tau, T]$ whenever there
exists a bit position $i \leq j$ for which $b_i \neq b'_i$.
\end{lemma}
{\em Proof:} This is a simple consequence of the orthogonality of
traditional pulse-position modulation. Since the underlying data bits
differ somewhere at or before $j$, then over each time-slot after
$(j-1)\tau$ the signals $x$ and $x'$ have disjoint support. 
\hfill $\Box$

\begin{figure}
\begin{center}
\setlength{\unitlength}{3947sp}%
\begingroup\makeatletter\ifx\SetFigFont\undefined%
\gdef\SetFigFont#1#2#3#4#5{%
  \reset@font\fontsize{#1}{#2pt}%
  \fontfamily{#3}\fontseries{#4}\fontshape{#5}%
  \selectfont}%
\fi\endgroup%
\begin{picture}(2140,1512)(1489,-1411)
\thinlines
{\color[rgb]{0,0,0}\put(1501,-1111){\framebox(300,1200){}}
}%
{\color[rgb]{0,0,0}\put(1501,-511){\line( 1, 0){300}}
}%
{\color[rgb]{0,0,0}\put(1951,-1111){\framebox(300,1200){}}
}%
{\color[rgb]{0,0,0}\put(1951,-511){\line( 1, 0){300}}
}%
{\color[rgb]{0,0,0}\put(1951,-811){\line( 1, 0){300}}
}%
{\color[rgb]{0,0,0}\put(1951,-211){\line( 1, 0){300}}
}%
{\color[rgb]{0,0,0}\put(2401,-1111){\framebox(300,1200){}}
}%
{\color[rgb]{0,0,0}\put(2401,-511){\line( 1, 0){300}}
}%
{\color[rgb]{0,0,0}\put(2401,-811){\line( 1, 0){300}}
}%
{\color[rgb]{0,0,0}\put(2401,-211){\line( 1, 0){300}}
}%
{\color[rgb]{0,0,0}\put(2401,-961){\line( 1, 0){300}}
}%
{\color[rgb]{0,0,0}\put(2401,-661){\line( 1, 0){300}}
}%
{\color[rgb]{0,0,0}\put(2401,-361){\line( 1, 0){300}}
}%
{\color[rgb]{0,0,0}\put(2401,-61){\line( 1, 0){300}}
}%
{\color[rgb]{0,0,0}\put(2851,-511){\line( 1, 0){300}}
}%
{\color[rgb]{0,0,0}\put(2851,-811){\line( 1, 0){300}}
}%
{\color[rgb]{0,0,0}\put(2851,-211){\line( 1, 0){300}}
}%
{\color[rgb]{0,0,0}\put(2851,-961){\line( 1, 0){300}}
}%
{\color[rgb]{0,0,0}\put(2851,-661){\line( 1, 0){300}}
}%
{\color[rgb]{0,0,0}\put(2851,-361){\line( 1, 0){300}}
}%
{\color[rgb]{0,0,0}\put(2851,-61){\line( 1, 0){300}}
}%
{\color[rgb]{0,0,0}\put(2851,-1036){\line( 1, 0){300}}
}%
{\color[rgb]{0,0,0}\put(2851,-886){\line( 1, 0){300}}
}%
{\color[rgb]{0,0,0}\put(2851,-736){\line( 1, 0){300}}
}%
{\color[rgb]{0,0,0}\put(2851,-511){\line( 1, 0){300}}
}%
{\color[rgb]{0,0,0}\put(2851,-586){\line( 1, 0){300}}
}%
{\color[rgb]{0,0,0}\put(2851,-436){\line( 1, 0){300}}
}%
{\color[rgb]{0,0,0}\put(2851,-286){\line( 1, 0){300}}
}%
{\color[rgb]{0,0,0}\put(2851,-136){\line( 1, 0){300}}
}%
{\color[rgb]{0,0,0}\put(2851,-1111){\framebox(300,1200){}}
}%
{\color[rgb]{0,0,0}\put(2851, 14){\line( 1, 0){300}}
}%
{\color[rgb]{0,0,0}\put(3301,-474){\line( 1, 0){300}}
}%
{\color[rgb]{0,0,0}\put(3301,-774){\line( 1, 0){300}}
}%
{\color[rgb]{0,0,0}\put(3301,-174){\line( 1, 0){300}}
}%
{\color[rgb]{0,0,0}\put(3301,-924){\line( 1, 0){300}}
}%
{\color[rgb]{0,0,0}\put(3301,-624){\line( 1, 0){300}}
}%
{\color[rgb]{0,0,0}\put(3301,-324){\line( 1, 0){300}}
}%
{\color[rgb]{0,0,0}\put(3301,-24){\line( 1, 0){300}}
}%
{\color[rgb]{0,0,0}\put(3301,-999){\line( 1, 0){300}}
}%
{\color[rgb]{0,0,0}\put(3301,-849){\line( 1, 0){300}}
}%
{\color[rgb]{0,0,0}\put(3301,-699){\line( 1, 0){300}}
}%
{\color[rgb]{0,0,0}\put(3301,-474){\line( 1, 0){300}}
}%
{\color[rgb]{0,0,0}\put(3301,-549){\line( 1, 0){300}}
}%
{\color[rgb]{0,0,0}\put(3301,-399){\line( 1, 0){300}}
}%
{\color[rgb]{0,0,0}\put(3301,-249){\line( 1, 0){300}}
}%
{\color[rgb]{0,0,0}\put(3301,-99){\line( 1, 0){300}}
}%
{\color[rgb]{0,0,0}\put(3301, 51){\line( 1, 0){300}}
}%
{\color[rgb]{0,0,0}\put(3309,-511){\line( 1, 0){300}}
}%
{\color[rgb]{0,0,0}\put(3309,-811){\line( 1, 0){300}}
}%
{\color[rgb]{0,0,0}\put(3309,-211){\line( 1, 0){300}}
}%
{\color[rgb]{0,0,0}\put(3309,-961){\line( 1, 0){300}}
}%
{\color[rgb]{0,0,0}\put(3309,-661){\line( 1, 0){300}}
}%
{\color[rgb]{0,0,0}\put(3309,-361){\line( 1, 0){300}}
}%
{\color[rgb]{0,0,0}\put(3309,-61){\line( 1, 0){300}}
}%
{\color[rgb]{0,0,0}\put(3309,-1036){\line( 1, 0){300}}
}%
{\color[rgb]{0,0,0}\put(3309,-886){\line( 1, 0){300}}
}%
{\color[rgb]{0,0,0}\put(3309,-736){\line( 1, 0){300}}
}%
{\color[rgb]{0,0,0}\put(3309,-511){\line( 1, 0){300}}
}%
{\color[rgb]{0,0,0}\put(3309,-586){\line( 1, 0){300}}
}%
{\color[rgb]{0,0,0}\put(3309,-436){\line( 1, 0){300}}
}%
{\color[rgb]{0,0,0}\put(3309,-286){\line( 1, 0){300}}
}%
{\color[rgb]{0,0,0}\put(3309,-136){\line( 1, 0){300}}
}%
{\color[rgb]{0,0,0}\put(3309,-1111){\framebox(300,1200){}}
}%
{\color[rgb]{0,0,0}\put(3309, 14){\line( 1, 0){300}}
}%
{\color[rgb]{0,0,0}\put(3317,-1074){\line( 1, 0){300}}
}%
\put(1651,-1411){\makebox(0,0)[b]{\smash{\SetFigFont{8}{10}{\rmdefault}{\mddefault}{\updefault}{\color[rgb]{0,0,0}1}%
}}}
\put(2101,-1411){\makebox(0,0)[b]{\smash{\SetFigFont{8}{10}{\rmdefault}{\mddefault}{\updefault}{\color[rgb]{0,0,0}2}%
}}}
\put(2551,-1411){\makebox(0,0)[b]{\smash{\SetFigFont{8}{10}{\rmdefault}{\mddefault}{\updefault}{\color[rgb]{0,0,0}3}%
}}}
\put(3001,-1411){\makebox(0,0)[b]{\smash{\SetFigFont{8}{10}{\rmdefault}{\mddefault}{\updefault}{\color[rgb]{0,0,0}4}%
}}}
\put(3451,-1411){\makebox(0,0)[b]{\smash{\SetFigFont{8}{10}{\rmdefault}{\mddefault}{\updefault}{\color[rgb]{0,0,0}5}%
}}}
\end{picture}
\end{center}
\caption{The sub-slots in each time-slot are refinements of the
 sub-slots in the previous time-slot. This gives rise to a natural tree
 structure and the semi-orthogonality property of the code.} 
\label{fig:refined_slots}
\end{figure}
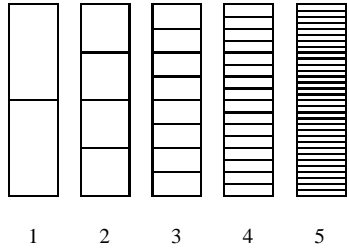

The time slot $[(k-1)\tau, k\tau]$ is divided into $2^k$ disjoint
sub-slots and all the energy ($E_b$) is put into the sub-slot that
corresponds to the realization of $B_1^k$ that has been seen so far at
the encoder. If the target rate is $R = \frac{1}{\tau}$ and the target
average power per unit time is $P$, then by using $E_b = \frac{P}{R}$,
it is clear that this scheme meets the target power constraint --- not
just when we average over the realizations of the incoming bits $B$,
but for every possible sequence of bits.
Figure~\ref{fig:refined_slots} illustrates the natural tree structure
of this code.

The decoder is assumed to have a target delay of $d \tau$ seconds and
to be interested in estimating the value of the bits with that delay.
In order to study asymptotic behavior, the interest is in the case of
$d$ large but finite.

\section{Analysis of achievable  $P_e$ with delay}

Because this is a sequential encoding scheme that is going to be used
with finite delay, the relevant error-event is a bit-error, not a
block error. The goal is for the probability of bit-error to go to
zero with delay. A code is considered to achieve reliability $E_a(R)$
if there exists a rate-dependent constant $K'$ so that
$$P(\hat{B}_i \neq B_i) \leq K' e^{-d\tau E_a(R)}$$
for every $i > 0, d > 0$. 

Our focus is on ML decoding. To get $\hat{B}_i$, the decoder has
access to the received waveform $Y(t)$ over the interval $t \in
[0,(i+d)\tau]$. Since no prior distribution over the $B_i$ is assumed,
the following ML strategy is used:
\begin{itemize}
 \item For every possible bit-sequence $\check{b}_1^{i+d}$, compute the
 log-likelihood $\ln p(Y([0,(i+d)\tau])=y([0,(i+d)\tau])|B_1^{i+d} =
 \check{b}_1^{i+d})$. By the white nature of the noise:
\begin{eqnarray} \label{eqn:whiteconsequence}
& \ln p(Y([0,(i+d)\tau])=y([0,(i+d)\tau])|B_1^{i+d} = \check{b}_1^{i+d})
\nonumber \\
& = \sum_{j=1}^{i+d} \ln
p_Y(y([(j-1)\tau,j\tau])|B_1^{j} = \check{b}_1^{j}) 
\end{eqnarray}

 \item Pick the most likely sequence $\tilde{B}_1^{i+d}$ and emit its
 $i$-th position. In the white-Gaussian case, this will reduce to a
 picking the bit-sequence that results in the minimum Euclidean
 distance between waveforms.
\end{itemize}

It is important to note that the decisions are not remembered in this
decoder. In principle, it recomputes the ML path each time. 

\subsection{Suffix-error analysis}
Suppose that a genie gave the decoder access to the correct value of
the bits $B_1^{i-1}$. Since it knows the truth before time $(i-1)\tau$
and (\ref{eqn:whiteconsequence}) tells us that the log-likelihood is
additive across time, the decoder only needs to consider $\ln
p(Y([(i-1)\tau,(i+d)\tau])=y([(i-1)\tau,(i+d)\tau])|B_1^{i+d} =
\check{b}_1^{i+d})$. The total duration of the relevant signal is thus
$(d+1)\tau$.

The only way an error can occur is if one of the $2^{d}$ bitstreams
with $\check{b}_i \neq b_i$ has a larger likelihood than the true
stream. By Lemma~\ref{lem:semiorthogonal}, the true waveform is
orthogonal to all the false waveforms under consideration. Recalling
that the block error probability analysis in Chap.~8 of
\cite{gallager} uses only the union bound and the fact that the true
waveform is orthogonal to each of the false ones\footnote{The
  analysis in \cite{gallager} proceeds by approximating the error
  event by the union of two events: that the noise in the direction of
  the true codeword is large and the event that a false codeword beats
  the true codeword conditioned on the fact that the noise in the
  direction of the true codeword is small. The standard union bound
  over all the false codewords is then used for the second event. By
  adjusting what is meant by ``large/small,'' the appropriate
  probabilities, the two terms are matched in the exponent and
  this gives the desired exponential bound.}, we can immediately apply
(\ref{eqn:proberrorblock}) to see that
\begin{equation} \label{eqn:suffix}
P(\hat{B}_i \neq B_i \mbox{ at delay }d | B_1^{i-1} \mbox{ known}) \leq K e^{-(d+1)\tau E_{orth}(R)}
\end{equation}

\subsection{Dealing with the uncertain prefix} \label{sec:prefix}

The actual decoder does not know $B_1^{i-1}$. However, the error
probability can be bounded as follows:
\begin{eqnarray} \label{eqn:prefixsum}
& & P(\hat{B}_i \neq B_i \mbox{ at delay }d) \nonumber\\
& \leq & 
\sum_{j=0}^{i-1} P(\hat{B}_{i-j} \neq B_{i-j} \mbox{ at delay }d+j |
B_1^{i-1-j} \mbox{ known})
\end{eqnarray}

since to make an error at bit $i$ the most likely sequence has to
differ from the true sequence at $i$ or earlier. The regular union
bound then gives us (\ref{eqn:prefixsum}) with the earlier
positions having correspondingly increased delays. 

Combining (\ref{eqn:prefixsum}) with (\ref{eqn:suffix}) gives us: 
\begin{eqnarray*}
& & P(\hat{B}_i \neq B_i \mbox{ with delay }d) \\
& \leq & \sum_{j=0}^{i-1} K e^{-(d+1+j)\tau E_{orth}(R)} \\
& < & K (\sum_{j=0}^{\infty} e^{-(j+1)\tau E_{orth}(R)}) e^{-d\tau E_{orth}(R)} \\
& = & \frac{K}{e^{\tau E_{orth}(R)} - 1} e^{-d\tau E_{orth}(R)}
\end{eqnarray*}
which gives the desired result with $E_a(R) = E_{orth}(R)$ and no
dependence on either the bit position $i$ and delay $d$. 

\begin{theorem} \label{thm:repeatedppm}
The repeated pulse position modulation code under maximum-likelihood
decoding for individual bits with delay $d\tau$ achieves the
orthogonal coding error exponent for every delay and bit position.
\begin{equation} \label{eqn:anytimeppm}
P(\hat{B}_i \neq B_i \mbox{ with delay }d) 
\leq K' e^{-d\tau E_{orth}(R)}
\end{equation}
\end{theorem}
\vspace{0.1in}

An immediate consequence of theorem \ref{thm:repeatedppm} is that this
code achieves zero probability of error on every bit in the limit of
large delays. The limit here is purely at the decoder rather than
being over encoder-decoder pairs. Consequently, it shows more clearly
what the nature of reliable communication can be over the
infinite-bandwidth channel. Using the language of anytime reliability
\cite{SahaiThesis, ControlPartI, OurSourceCodingPaper},
Theorem~\ref{thm:repeatedppm} establishes a lower bound on the anytime
reliability of the infinite-bandwidth AWGN channel without feedback.

The code as described is a pulse-position modulation variation that
ends up requiring unboundedly large peak amplitudes while meeting a
hard power constraint on the $\tau$ timescale. Because all that is
required is orthogonality on each time slot, any other orthogonal
signaling could be used. In particular, the code could use signals
that have constant amplitude and just change abruptly in phase. 

\section{Upper bound on the error exponent with delay}

In \cite{PinskerNoFeedback}, Pinsker gave an derivation for the BSC
showing that non-block codes without feedback could not beat the
Sphere-Packing bound for how fast the probability of bit-error decays
with end-to-end delay. This argument was given a more
careful\footnote{Pinsker in \cite{PinskerNoFeedback} claimed that the
  argument extended to cases when feedback was present which is
  false.} exposition and generalized to all DMCs without feedback in
\cite{OurUpperBoundPaper} with the Haroutunian bound
\cite{Haroutunian} playing\footnote{The Haroutunian bound is the same
  as the Sphere-Packing bound for symmetric channels and Pinsker's
  result is recovered.} the role of the Sphere-Packing bound.

Here, we quickly show how to generalize Wyner's block-coding converse
result from \cite{WynerAWGN} to the delay setting. Since Wyner's
argument was very block-specific (involving solid-angles, etc.), this
is done by extending the argument of Section~IV of
\cite{OurUpperBoundPaper} to the continuous-time AWGN case with a hard
amplitude constraint\footnote{At the expense of slightly more
  cumbersome notation, the argument easily generalizes to the hard
  ``average'' power-constraint model here where the energy emitted
  every $\tau$ seconds is limited to a hard constraint.}. Let us
examine the core steps of the proof from \cite{OurUpperBoundPaper}:
\begin{enumerate}
\item Use a rate $R$ code with fixed-delay $d\tau$ to construct a
  hypothetical long block-code with rate $R-\delta$: this step is
  unchanged.

\item Consider ``feedforward decoders'' that have genie access to the
  truth about past bits. This step is also unchanged and so the
  error-event on a bit can be restricted to what the channel is like
  for the $d\tau$ time steps between when the bit entered the encoder
  and when its estimate emerges from the decoder.

\item Run the code over a ``nearby'' channel whose capacity is only
  sufficient to sustain a rate of $R-2\delta$. The data-processing
  inequality then forces the bit-error process to have an entropy rate
  of at least $\delta$. This step is unchanged in spirit, but the key
  here is not to consider a channel with increased
  noise\footnote{Increasing the noise-intensity would cause trouble
    when trying to change measures back to the original channel since
    there are an exponential number of degrees of freedom --- this
    would give rise to a doubly-exponential bound on how the
    probability of error decays to zero.} but rather a channel that
  attenuates the input before adding the same noise as before.  The
  attenuation factor $\sqrt{\frac{R-2\delta}{C_{\infty}}}$ chosen is such
  that the post-attenuation power can barely sustain a rate of
  $R-2\delta$ and thus cannot sustain $R-\delta$.

\item Fano's inequality applied to the bit-error sequence reveals that
  the probability of bit-error under this channel's induced
  probability measure is at least $\delta'$. This step is also
  unchanged.

\item The feedforward nature of the decoder tells us that the error
  events only depend on the white noise realization for a duration of
  $d\tau$. This is also unchanged from \cite{OurUpperBoundPaper}.
  Here, Wyner's argument from \cite{WynerAWGN} also enters and it is
  clear that since the decoder knows the true bits from the past,
  there are only $2^{d}$ possible waveforms that could have been
  transmitted. Expanding the white noise in the appropriately aligned
  basis means that only $2^d$ dimensions of noise are relevant.
  Without loss of generality, assume that one of these dimensions is
  exactly aligned with the waveform that was actually transmitted
  during this duration.

\item The probability of the error event has to be evaluated under the
  true channel law rather than the ``nearby'' law. Here, the story is
  considerably {\em simpler} than the DMC case. This is because the
  ``nearby'' channel behaves exactly the same as the original channel
  in the $2^d - 1$ directions other than the true direction. Within
  the one true direction, typicality dictates that the nearby-channel
  noise is certainly going to be within $\pm Q^{-1}(\delta')
  \sqrt{N_0}$ where $Q^{-1}$ just appropriately inverts the CDF for a
  standard Gaussian. 
\end{enumerate}
Thus, under the original channel law, the noise only has to push the
$\sqrt{d\tau P}$ signal down to around: 
\begin{eqnarray*}
& & \sqrt{d\tau P} (\sqrt{\frac{R-2\delta}{C_{\infty}}}) \pm Q^{-1}(\delta')
  \sqrt{N_0} \\
& = & \sqrt{\frac{d\tau P (R-2\delta)}{(P/N_0)\log_2 e}} \pm Q^{-1}(\delta')
  \sqrt{N_0} \\
& = & \sqrt{\frac{d\tau (R-2\delta)N_0}{\log_2 e}} \pm Q^{-1}(\delta')
  \sqrt{N_0} \\
& = & \sqrt{N_0}(\sqrt{\frac{d\tau (R-2\delta)}{\log_2 e}} \pm
Q^{-1}(\delta')
\end{eqnarray*}
to cause an error. This means a noise for the original
channel\footnote{This is essentially what is going on in equation (29)
  of \cite{WynerAWGN}.} in this direction of around
\begin{eqnarray*}
&   & -\sqrt{d\tau P} + \sqrt{N_0}(\sqrt{\frac{d\tau (R-2\delta)}{\log_2 e}} \pm
Q^{-1}(\delta')  \\
& = & -\sqrt{\frac{d\tau C_\infty N_0}{\log_2 e}} + \sqrt{N_0}(\sqrt{\frac{d\tau (R-2\delta)}{\log_2 e}} \pm
Q^{-1}(\delta')  \\
& = & -\sqrt{d \tau N_0 \ln 2}(\sqrt{C_\infty} - \sqrt{R-2\delta} \pm
\frac{Q^{-1}(\delta')}{d \tau \ln 2})
\end{eqnarray*}
Thus the probability of error under the original channel law is at
least
\begin{eqnarray*}
\frac{\delta'}{2} \exp(-\frac{(\sqrt{d \tau N_0 \ln 2}(\sqrt{C_\infty}
  - \sqrt{R-2\delta} + \frac{Q^{-1}(\delta')}{d \tau \ln 2}))^2}{N_0})
& =  &
\frac{\delta'}{2} \exp(-d\tau \ln 2 (\sqrt{C_\infty} -
\sqrt{R-2\delta} + \frac{Q^{-1}(\delta')}{d \tau \ln 2})^2 
\end{eqnarray*}
Since $\frac{Q^{-1}(\delta')}{d \tau \ln 2}$ is as small as desired
when $d$ is large enough and $\delta,\delta'$ were arbitrary, this
essentially shows that the high rate expression for $E_{orth}(R)$ in
(\ref{eqn:orthogonalexponent}) cannot be beaten with delay. Thus the
semi-orthogonal code has essentially asymptotically optimal
performance with delay.

\section{Interpretations and extensions}
Is this a practical tool for data communication? It seems likely that
the answer to this question is negative. Orthogonal signaling can be
very wasteful of bandwidth and the semi-orthogonal scheme given here
actually uses $\infty$ bandwidth which is never available in practice.
The goal here is rather to refine our understanding of the role of
delay in reliable communication and the tradeoffs possible between the
encoder and decoder. It provides another extreme point balancing
block-codes on the other side. Consequently, it is best to consider
its theoretical rather than practical implications.

\subsection{Feedback}

Without feedback, running an infinite-constraint-length random
convolutional code would require performing an growing number of
computations per unit time at the encoder just to generate the next
channel input. Section II.C of \cite{OurUpperBoundPaper} gives a way
to use perfect tentative decision feedback to allow
infinite-constraint-length convolutional codes to run using only
bounded expected computation per time. Rather than convolving with the
data from time zero till now to generate the channel inputs, the idea
is to convolve against the current error sequence instead. The error
sequence and the data bits are in one-to-one correspondence given
knowledge of the current tentative estimates. Thus all the distance
properties of the random infinite-constraint-length code are preserved
along with all the expected probabilities of error. The encoding
complexity is reduced dramatically since all bit-estimates from the
distant past are almost always correct due to the exponential
convergence of bit-estimates to true values.

In the infinite-bandwidth case here, this same trick can be used to
reduce the bandwidth consumption in some similar ``expected'' sense.
Consider the family of orthogonal waveforms from
Figure~\ref{fig:blockwaveforms}. If both sine and cosine signals are
used, then waveform $i$ has a frequency of about $\frac{i}{2\tau}$.
Instead of transmitting a packet of energy $E_b$ that
corresponds to all the bits sent so far, the packet can correspond to
the current error-signal if the tentative decoder decisions are
available at the transmitter. Once again, this does not change the
semi-orthogonality property of the code because conditioned on the
known tentative decisions, all suffixes in the codebook are still
orthogonal to each other.

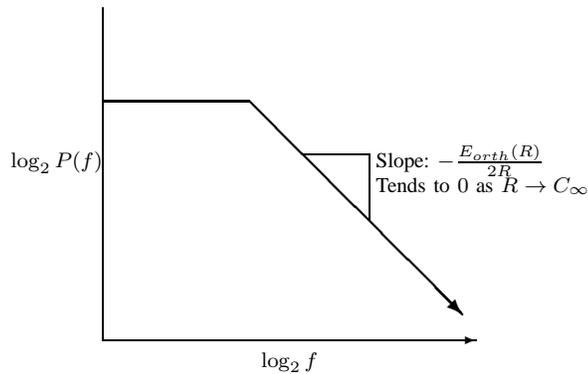
\begin{figure}
\begin{center}
\setlength{\unitlength}{2200sp}%
\begingroup\makeatletter\ifx\SetFigFont\undefined%
\gdef\SetFigFont#1#2#3#4#5{%
  \reset@font\fontsize{#1}{#2pt}%
  \fontfamily{#3}\fontseries{#4}\fontshape{#5}%
  \selectfont}%
\fi\endgroup%
\begin{picture}(4662,4120)(751,-3719)
\thinlines
{\color[rgb]{0,0,0}\put(3451,-1261){\line( 1, 0){750}}
\put(4201,-1261){\line( 0,-1){750}}
}%
\thicklines
{\color[rgb]{0,0,0}\put(1201,-661){\line( 1, 0){1650}}
\put(2851,-661){\vector( 1,-1){2400}}
}%
\thinlines
{\color[rgb]{0,0,0}\put(1201,389){\line( 0,-1){3750}}
\put(1201,-3361){\vector( 1, 0){4200}}
}%
\put(3301,-3661){\makebox(0,0)[b]{\smash{\SetFigFont{8}{14.4}{\rmdefault}{\mddefault}{\updefault}{\color[rgb]{0,0,0}$\log_2 f$}%
}}}
\put(680,-1411){\makebox(0,0)[b]{\smash{\SetFigFont{8}{14.4}{\rmdefault}{\mddefault}{\updefault}{\color[rgb]{0,0,0}$\log_2 P(f)$}%
}}}
\put(4276,-1411){\makebox(0,0)[lb]{\smash{\SetFigFont{8}{14.4}{\rmdefault}{\mddefault}{\updefault}{\color[rgb]{0,0,0}Slope: $-\frac{E_{orth}(R)}{2R}$}%
}}}
\put(4276,-1686){\makebox(0,0)[lb]{\smash{\SetFigFont{8}{14.4}{\rmdefault}{\mddefault}{\updefault}{\color[rgb]{0,0,0}Tends to $0$ as $R \rightarrow C_{\infty}$}%
}}}
\end{picture}
\end{center}
\caption{If tentative decision feedback is available, then
  ridiculously high frequencies are used only rarely as time goes
  on. Most of the time, the code will use low frequencies. The slope
  of the curve depends on the data rate: the higher the data rate, the
  more often high frequencies end up getting used.}
\label{fig:finitebandwidth}
\end{figure}

Because of the exponential convergence of the probability of bit
error, most of the early bits are correct and the errors are
concentrated in the most recent bits. This gives rise to a bandwidth
usage that is depicted in Figure~\ref{fig:finitebandwidth}. If the
earliest current error is $d$ bits ago, then a frequency of at most
$2^{\frac{d}{2}}$ needs to be used during this transmission. However
the probability of that scales proportional to
$2^{-E_{orth}(R)d}$. The two exponents fight each other to a
power-law. 

Note, while this seems to make the bandwidth requirements more
reasonable, it does not make this a practical coding idea. The rolloff
at high frequencies in traditional communication systems is not just a
matter of not causing undue interference to users in adjacent bands.
It is also about making the code itself robust to the behavior of
users in those adjacent bands. Even when its transmissions are kept to
low frequencies, the decoder remains very sensitive to the noise at
all frequencies.

\subsection{Capacity per unit cost} 

Verdu's capacity per unit cost framework \cite{verducost} is the
natural generalization of the infinite-bandwidth power-constrained
AWGN channel. To see how the main result of this paper translates, the
error calculation can be reinterpreted as a function of the number of
bits intervening rather than in terms of the time delay and rate.
Apply the substitutions $E_b = \frac{P}{R}$, $\tau = \frac{1}{R}$ to
(\ref{eqn:awgncapacityintime}), (\ref{eqn:anytimeppm}), and
(\ref{eqn:orthogonalexponent}) to get:

\begin{equation} \label{eqn:anytimeppmeb}
P(\hat{B}_i \neq B_i \mbox{ with }d\mbox{ bits intervening}) \leq K' e^{-d E_{orth}(\frac{E_b}{N_0})}
\end{equation}
where $E_{orth}(\frac{E_b}{N_0}) =$
\begin{equation} \label{eqn:orthogonalexponenteb}
\left\{ \begin{array}{rl} 
                      (\frac{E_b}{N_0}(\frac{1}{2\ln 2}) - 1)\ln 2 &
                      \mbox{if } \frac{E_b}{N_0} > 4 \ln 2 \\
                      (\sqrt{\frac{E_b}{N_0}(\frac{1}{\ln 2})} - 1)^2
                      \ln 2 & \mbox{if }\ln 2 < \frac{E_b}{N_0} \leq 4 \ln 2 \\ 
                      0 & \mbox{otherwise} \end{array}\right.
\end{equation}

With this done, the infinite-bandwidth assumption can be dropped as
long as the interest is only in the probability of error going to zero
as a function of intervening energy. This gives rise to the natural
``sequential'' version of Verdu's capacity per unit cost framework:

\begin{itemize}
 \item There is a zero-cost channel input available. 

 \item The encoder gets access to the bits one at a time and can use
       any number of channel uses it likes. 

 \item The total cost of the channel uses so far must be less than
       $E_b$ times the number of bits the encoder has received.

 \item The decoder wants to estimate the values of all the bits, but
       is willing to wait until the encoder has spent a certain extra
       amount $d E_b$ more than it had when it first got access to the
       desired bit.
\end{itemize}

As the encoding strategy in \cite{verducost} is a kind of pulse
position modulation, the repeated pulse-position-modulation strategy
described here translates directly into that framework.\footnote{We
  will use the zero-cost channel input in most places, except for the
  position that corresponds to all the bits so far. There we use the
  appropriate more expensive channel input.} Interpret the sub-slots
depicted in figure \ref{fig:many_bits} as individual channel uses. The
semi-orthogonality property of Lemma~\ref{lem:semiorthogonal}
translates directly into a semi-disjoint support property. If the ML
detector in the natural form is used, then
(\ref{eqn:whiteconsequence}) continues to apply as appropriately
interpreted. For error analysis analogous to (\ref{eqn:suffix}), the
ML performance can be bounded by the suboptimal, yet simple,
hypothesis-testing based decoder in \cite{verducost}. Since the error
analysis in \cite{verducost} relies only on the disjointness of the
true codeword to each false one individually, all the arguments given
here extend directly.\footnote{For a given prefix, the true suffix has
  $d$ expensive channel inputs at the appropriate places.  The false
  suffices all have the zero cost channel input in that place.
  Similarly, they all have their expensive channel inputs where the
  true suffix has zeros. Thus the pairwise competition works.} All
that is required for the prefix-argument of Section~\ref{sec:prefix}
to work is that the probability of error go to zero exponentially in
$d$. Although this is not explicit in the stated proof in
\cite{verducost}, it does indeed hold.\footnote{Look at the set of
  types that are accepted as representing that a ``pulse'' is present
  at the appropriate set of channel inputs. The threshold corresponds
  to how much of a margin around the type $P_0$ we are going to
  accept. All that matters is that there is a margin and so the
  probability of missing the true ``pulse'' is exponentially small
  with $d$. The Stein's lemma argument already gives us an
  exponentially small probability of false alarm using the union
  bound. By choosing the threshold to equalize the exponents, a single
  exponent is obtained. This decoder might not give the best possible
  exponent, but all that is needed here is that it give some nonzero
  exponent. To see why is it not the best, the reader is encouraged to
  carry out this analysis for the AWGN case and compare to
  $E_{orth}$.}

There is only one new consideration: integer effects. In the
continuous amplitude situation, the transmitter could measure out any
desired energy and pour it into a disjoint time period. For a DMC with
an input cost constraint, there might be no way to hit the desired
cost with a single input and so some way of time-sharing between
inputs is essential.

It is here that we can see a role for a finite additional ``delay''
being imposed at the encoder. If the encoder decides to ``burst'' its
output, it can do so by buffering up bits (spending no channel cost)
until it has $L$ of them ready to go, and is willing to spend $L E_b$
cost to do the incremental encoding. By letting $L$ get large, $L E_b$
gets big enough to smooth out any integer constraint.\footnote{Having
a budget of $L E_b$ allows us to choose the appropriate mix of channel
inputs in a sub-slot that together cost less than $L E_b$ but have
close to the desired divergence.} However, the probability of error
analysis given earlier continues to hold and the probability of
$L$-burst error will go to zero exponentially at the correct rate with
respect to cost increments.

In the capacity per unit cost formulation, this scheme or finitely
truncated versions of it might actually have practical
consequences. Consider a sensor network with very little energy
available for long-range communication, but also very little data to
send. If the data is going to be used by some application, the sensor
may not know what the acceptable ``delay'' is. By using an anytime or
delay-universal scheme like the one presented here, a sensor might be
able to leave that choice to the decoder.

\section{Conclusions and open problems}
This correspondence shows how to communicate over the
infinite-bandwidth AWGN channel in the non-block setting where
bit-errors and end-to-end delays are important. An explicit
semi-orthogonal coding scheme was developed and analyzed to show that
the block-coding $E_{orth}(R)$ exponents are achievable with
end-to-end delay. In the high-rate regime, a tight upper bound was
given showing that the code is essentially optimal. While the basic
semi-orthogonal coding strategy uses an increasing amount of bandwidth
with time, if noiseless tentative decision feedback is available, the
bandwidth usage can be made softer with high frequencies used only
rarely. 

The ideas here are taken only one-step toward the
capacity-per-unit-cost formulation. It remains open to see what the
optimal error exponents are with incremental cost and how to achieve
them. In addition, the semi-orthogonal coding strategy for the
capacity-per-unit-cost problem has a time-rate that tends to zero very
rapidly as time advances. It is not clear if tricks similar to this
paper can be used to combat this problem even if tentative decision
feedback was available at the transmitter.

\section*{Acknowledgments}

Thanks to Pramod Viswanath for suggesting that there might be a
connection to the capacity per unit cost formulation. Thanks also go
to the Berkeley students in the digital communications course in 2003
whose questions helped push the author to develop this material as a
teaching aid.

\bibliographystyle{IEEEtran}
\bibliography{IEEEabrv,./MyMainBibliography}

\begin{thebibliography}{10}
\providecommand{\url}[1]{#1}
\csname url@rmstyle\endcsname
\providecommand{\newblock}{\relax}
\providecommand{\bibinfo}[2]{#2}
\providecommand\BIBentrySTDinterwordspacing{\spaceskip=0pt\relax}
\providecommand\BIBentryALTinterwordstretchfactor{4}
\providecommand\BIBentryALTinterwordspacing{\spaceskip=\fontdimen2\font plus
\BIBentryALTinterwordstretchfactor\fontdimen3\font minus
  \fontdimen4\font\relax}
\providecommand\BIBforeignlanguage[2]{{%
\expandafter\ifx\csname l@#1\endcsname\relax
\typeout{** WARNING: IEEEtran.bst: No hyphenation pattern has been}%
\typeout{** loaded for the language `#1'. Using the pattern for}%
\typeout{** the default language instead.}%
\else
\language=\csname l@#1\endcsname
\fi
#2}}

\bibitem{WolfLecture}
J.~Wolf, ``An introduction to turbo codes: The quintessential channel coding
  technique,'' Tech. Rep., Mar. 2000.

\bibitem{gallager}
R.~G. Gallager, \emph{Information Theory and Reliable Communication}.\hskip 1em
  plus 0.5em minus 0.4em\relax New York, NY: John Wiley, 1971.

\bibitem{LiuViswanath}
T.~Liu and P.~Viswanath, ``Opportunistic orthogonal writing on dirty paper,''
  \emph{{IEEE} Trans. Inform. Theory}, vol.~52, no.~5, pp. 1828--1846, May
  2006.

\bibitem{WynerAWGN}
A.~D. Wyner, ``On the probability of error for communication in white
  {G}aussian noise,'' \emph{{IEEE} Trans. Inform. Theory}, vol.~13, no.~1, pp.
  86--90, Jan. 1967.

\bibitem{ForneyML}
G.~D. Forney, ``Convolutional codes {II}. maximum-likelihood decoding,''
  \emph{Information and Control}, vol.~25, no.~3, pp. 222--266, July 1974.

\bibitem{PinskerNoFeedback}
M.~S. Pinsker, ``Bounds on the probability and of the number of correctable
  errors for nonblock codes,'' \emph{Problemy Peredachi Informatsii}, vol.~3,
  no.~4, pp. 44--55, Oct./Dec. 1967.

\bibitem{OurUpperBoundPaper}
\BIBentryALTinterwordspacing
A.~Sahai, ``Why block length and delay are not the same thing,'' \emph{{IEEE}
  Trans. Inform. Theory}, submitted. [Online]. Available:
  \url{http://www.eecs.berkeley.edu/\~\,$\!$sahai/Papers/FocusingBound.pdf}
\BIBentrySTDinterwordspacing

\bibitem{JelinekSequential}
F.~Jelinek, ``Upper bounds on sequential decoding performance parameters,''
  \emph{{IEEE} Trans. Inform. Theory}, vol.~20, no.~2, pp. 227--239, Mar. 1974.

\bibitem{ForneySeq}
G.~D. Forney, ``Convolutional codes {III}. sequential decoding,''
  \emph{Information and Control}, vol.~25, no.~3, pp. 267--297, July 1974.

\bibitem{SahaiThesis}
A.~Sahai, ``Any-time information theory,'' Ph.D. dissertation, Massachusetts
  Institute of Technology, Cambridge, MA, 2001.

\bibitem{ControlPartI}
A.~Sahai and S.~K. Mitter, ``The necessity and sufficiency of anytime capacity
  for stabilization of a linear system over a noisy communication link. part
  {I}: scalar systems,'' \emph{{IEEE} Trans. Inform. Theory}, vol.~52, no.~8,
  pp. 3369--3395, Aug. 2006.

\bibitem{OurSourceCodingPaper}
\BIBentryALTinterwordspacing
------, ``Source coding and channel requirements for unstable processes,''
  \emph{{IEEE} Trans. Inform. Theory}, Submitted, 2006. [Online]. Available:
  \url{http://www.eecs.berkeley.edu/\~\,$\!$sahai/Papers/anytime.pdf}
\BIBentrySTDinterwordspacing

\bibitem{Haroutunian}
E.~A. Haroutunian, ``Lower bound for error probability in channels with
  feedback,'' \emph{Problemy Peredachi Informatsii}, vol.~13, no.~2, pp.
  36--44, 1977.

\bibitem{verducost}
S.~Verd\'{u}, ``On channel capacity per unit cost,'' \emph{{IEEE} Trans.
  Inform. Theory}, vol.~36, no.~9, pp. 1019--1030, Sept. 1990.

\end{thebibliography}
\end{document}